\documentclass[aps,preprint,amsmath,amssymb]{revtex4}

\usepackage{graphicx}

\begin{document}
\title{\Large \bf Lepton flavor violation in $\tau$ decays }
\date{\today}
\author{\large \bf
Chuan-Hung~Chen$^{1,3}$\footnote{Email: physchen@mail.ncku.edu.tw}
and Chao-Qiang~Geng$^{2,3}$\footnote{Email: geng@phys.nthu.edu.tw} }
\affiliation{
$^{1}$Department of Physics, National Cheng-Kung University, Tainan, 701 Taiwan\\
$^{2}$Department of Physics, National Tsing-Hua University, Hsin-Chu
, 300 Taiwan  \\
$^{3}$National Center for Theoretical Sciences, Taiwan }

\begin{abstract}
We study the lepton flavor violation (LFV) in tau decays in the
framework of the supersymmetric seesaw mechanism with
nonholomorphic terms for the lepton sector at a large
$\tan\beta$.
In particular, we analyze two new decay
modes $\tau\to \ell f_0(980)$ and $\tau\to \ell K^{+} K^{-}$
arising from the scalar boson exchanges contrast to
 $\tau\to \ell\eta^{(\prime)}$ from the pseudoscalar ones.
We find that the decay branching
ratios of the two new modes could be not only as large as the current
upper limits of $O(10^{-7})$, but also larger than those of $\tau\to
\ell \eta^{(\prime)}$.
Experimental searches for the two modes are important for the LFV
induced by the scalar-mediated mechanism.
In addition, we show that the decay
branching ratios of $\tau\to \ell \mu^{+} \mu^{-}$ are related to
those of $\tau\to \ell \eta$ and $\tau\to \ell f_0(980)$.

\end{abstract}
\maketitle

In the standard model (SM), since the neutrinos are regarded as
massless particles, the processes associated with lepton flavors are
always conserved. Inspired by the discoveries of nonzero neutrino
masses \cite{atom,solar}, it has been studied enormously how to
generate the neutrino masses which are less than a few $eV$.
By supplementing
with singlet right-handed Majorana neutrinos with masses $M_R$
required to be around the scale of unified theory,
it is found that the seesaw mechanism is one of natural ways
\cite{seesaw} to obtain the small neutrino masses. Accordingly, in
non-SUSY models, it is easy to understand that the
effects of
the lepton flavor violation (LFV)
  are suppressed by $1/M_{R}$. However, in
 models with SUSY, due to the nondiagonal neutrino mass matrix,
 the flavor conservation in the slepton sector at
the unified scale will be violated at the $M_{R}$ scale via
renormalization \cite{BM,HMTY,BK_PRL}. The flavor violating effects
could propagate to the electroweak scale so that instead of
$1/M_{R}$, the suppression of the LFV could be $1/M_{SUSY}$ with
$M_{SUSY}\sim O(TeV)$ being the typical mass of the SUSY particle.
 Consequently, the lepton flavor violating processes, such as
$\ell^{-}_{i}\to \ell^{-}_{j} \gamma$ and $\ell^{-}_{i}\to
\ell^{-}_{j} \ell^{+}_{k} \ell^{-}_{k}$, become detectible at the
low energy scale.
The LFV has been extensively studied in the literature. For examples,
 $\tau\to \mu \eta$, $B\to (e,\; \mu) \tau$  and
 the $\mu - (e,\tau)$ conversions can be found in
Refs.~\cite{LFV1,LFV2,MORE,mu-e,mu-ePara,mu-tau},
while that to the detection of the LFV in colliders is given in
Ref.~\cite{LFV_collider}.

In the large $\tan\beta$ region, it has been pointed out that the
nonholomorphic Yukawa interactions \cite{Hamzaoui:1998nu,qf1,qf2,IR}
play very important roles for flavor changing neutral currents
(FCNCs) in the quark sector. In the SUSY-seesaw model, the
nonholomorphic terms \cite{BK_PRL} in the lepton sector naturally
induce the LFV due to the Higgs couplings.
It has been shown that the contribution to
the decay of $\tau\to 3\mu$ from the Higgs-mediated
LFV at large $\tan\beta$ could be much larger than that from
 $\tau\to \mu \gamma\to \mu\mu^{+}\mu^{-}$  \cite{BK_PRL,Parry}.
 Recently, the experimental limits on the radiative decays of
$\tau\to \ell
\gamma$ ($\ell=e,\; \mu$) have been improved from $O(10^{-6})$
\cite{PDG04} to $O(10^{-7})$ \cite{belle-tau,babar-tau}. Moreover,
the sensitivity of probing the LFV in $\tau$ decays with
 single pseudoscalar $(P)$ or vector $(V)$ and double mesons
 in the final states,
$i.e.$, $\tau\to \ell (P,\; V)$ and  $\tau\to \ell PP$,
 have also reached $O(10^{-7})$ \cite{belle-tau-hh}.
In this paper,
we will simultaneously analyze
 $\tau\to \ell \gamma$ and $\tau\to\ell X $,
 where $X$ are $\mu^{+} \mu^{-}$, $\eta^{(\prime)}$, $\phi$,
$f_{0}(980)$, $\sigma(600)$ and $K^{+} K^{-}$, respectively,
in the Higgs-mediated mechanism.
In particular, we would like to check whether it is possible to have large  rates for the processes beside the mode of $\tau\to 3\mu$.
 Note that the decays of
$\tau\to \ell S$ with $S=f_{0}(980)$ and $\sigma(600)$ and
$\tau\to \ell (\phi, K^{+}K^{-})$
have not been
explored previously based on the Higgs-mediated mechanism in the literature, while
 $\tau\to \ell P$
 have been studied in Refs. \cite{LFV1,LYZ,MORE}.
%

We start with the Higgs-mediated mechanism.
It is known that by the induced slepton flavor mixing, the effective
Lagrangian with induced nonholomophic terms for the Higgs bosons
coupling to
leptons is given by \cite{BK_PRL}
\begin{eqnarray}
  -{\cal L}_{\rm eff}&=&\bar{E}_{Ri} Y_{ i} \left[
  \delta_{ij} H^{0}_{d}  + \left( \epsilon_{1} \delta_{ij} +\epsilon_{2} I_{ij} \right) H^{0*}_{u}\right]
   E_{Lj}  +h.c.\,, \nonumber \\
   &=& \bar{E}_{R} M^{0}_{\ell} E_{L} +h.c.\,, \label{eq:mass}
\end{eqnarray}
where $Y$ denotes the diagonalized Yukawa matrix of leptons,
$I_{ij}=(\Delta m^2_{\tilde{L}})_{ij}/m^2_{0}$ and $\epsilon_{1(2)}$
is related to the induced lepton flavor conserving
(violating) effect, expressed by \cite{BK_PRL}
 \begin{eqnarray}
  \epsilon_{1}&\simeq &  \frac{\alpha_{1}}{8\pi} \mu M_{1}
  \left[ 2f_{1}(M^2_{1}, m^2_{\tilde{\ell}_{L}}, m^2_{\tilde{\ell}_{R}})
  - f_{1}(M^2_{1}, \mu^2, m^2_{\tilde{\ell}_{L}})
  +2 f_{1}(M^2_{1}, m^2_{\tilde{\ell}_{L}}, m^2_{\tilde{\ell}_{R}})\right]\nonumber \\
  &+& \frac{\alpha_{2}}{8\pi} \mu M_{2} \left[ f_{1}(\mu^2, m^2_{\tilde{\ell}_{L}}
  + 2 f_{1}(\mu^2, m^2_{\tilde{\nu}_{\ell}}, M^2_{2})\right]\, , \nonumber \\
  \epsilon_{2}& \simeq& \frac{\alpha_{1}}{8\pi} \mu M_{1} m^2_{0}
  \left[ 2f_{2}(M^2_{1}, m^2_{\tilde{\ell}_{L}},m^2_{\tilde{\tau}_{L}}, m^2_{\tilde{\tau}_{R}})
  - f_{2}(\mu^2, m^2_{\tilde{\ell}_{L}}, m^2_{\tilde{\tau}_{L}},M^2_{1})
  \right] \nonumber \\
  &+& \frac{\alpha_{2}}{8\pi} \mu M_{2} m^2_{0}
  \left[ f_{2}(\mu^2, m^2_{\tilde{\ell}_{L}},
  m^2_{\tilde{\tau}_{L}},M^2_{2})+2 f_{2}(\mu^2, m^2_{\tilde{\nu}_{\ell}},
  m^2_{\tilde{\nu}_{\tau}},M^2_{2}) \right]\,, \label{eq:epsilon}
 \end{eqnarray}
where
$M_{1,2}$ are the masses of gauginos from the soft
SUSY breaking terms, $\mu$ stands for the mixing of $H_{u}$ and
$H_{d}$,
 \begin{eqnarray*}
   f_1(x,y,z)&=& - { xy\ln(x/y)+yz\ln(y/z)+zx\ln(z/x)\over (x-y)(y-z)(z-x)}\,,\nonumber \\
   f_2(w,x,y,z)&=& - { w\ln w \over (w-x)(w-y)(w-z)}- {\rm cyclic},
 \end{eqnarray*}
and  $\alpha_{1(2)}=g^2_{1(2)}/4\pi$ with $g_{1(2)}$ corresponding to the gauge coupling of the $U(1)(SU(2))$
symmetry.
Due to the nonholomorphic term $\epsilon_2 E_{ij}$, the lepton
mass matrix
  is not diagonal anymore. Consequently, after
rediagonalizing the lepton mass matrix, the lepton flavor changing
neutral interactions through the Higgs bosons appear. Since the
nonholomorphic terms are expected to be much less than unity, to
obtain the LFV, we take the unitary matrices used for diagonalizing
lepton mass matrix to be $U_{L(R)}\approx {\bf 1}+\Delta_{L(R)}$ as
a leading expansion of $\epsilon_{2} E_{ij}$, where $\Delta_{L(R)}$
are $3\times 3$ matrices. From Eq.~(\ref{eq:smass}) and
$(\Delta m^2_{\tilde{L}})_{ij}=(\Delta m^2_{\tilde{L}})_{ji}$, we
may set $\Delta_{L}=\Delta_{R}=\Delta$. Hence, the diagonal mass
matrix in Eq.~(\ref{eq:mass}) could be obtained by
 \begin{eqnarray*}
  U M ^{0}_{\ell} U^{\dagger}\approx ({\bf 1}+\Delta) M^{0}_{\ell} ({\bf 1} - \Delta)=M^{dia}_{\ell}\, ,
  \label{eq:mass-dia}
 \end{eqnarray*}
where $M^{dia}_{\ell}$ is the physical mass matrix of the lepton
with the diagonal elements being
$(M^{dia}_{\ell})_{ii}=(m_{e},\, m_{\mu},\, m_{\tau})$. At the
leading order, we get
  \begin{eqnarray*}
  \left(M^{0}_{\ell}\right)_{ii} \approx  \left(M^{dia}_{\ell } \right)_{ii}\,,\
   \Delta_{ij}\approx \frac{(M^{0}_{\ell})_{ij}}{(M^{0}_{\ell})_{ii}-(M^{0}_{\ell})_{jj}}  \ \ \ (\rm i\neq j).
  \end{eqnarray*}
In terms of the physical mass eigenstates of the Higgs bosons, represented by \cite{Higgs}
  \begin{eqnarray*}
   ReH^{0}_{d}&=&v_{d} + \frac{1}{\sqrt{2}} \left[ \cos\alpha H^{0} - \sin\alpha h^{0}\right],\
   ReH^{0}_{u}=v_{u} + \frac{1}{\sqrt{2}} \left[ \sin\alpha H^{0} + \cos\alpha h^{0}\right],
   \nonumber \\
   ImH^{0}_{d}&=& \frac{1}{\sqrt{2}}\left[\cos\beta G^{0} - \sin\beta A^{0} \right],\
%
      ImH^{0}_{u}= \frac{1}{\sqrt{2}}\left[\sin\beta G^{0} + \cos\beta A^{0} \right],
  \end{eqnarray*}
where $\alpha$ is the mixing angle of the two CP-even neutral
scalars, the interactions for the LFV via the Higgs-mediated
mechanism are expressed by
 \begin{eqnarray}
  {\cal H}^{i\neq j}_{\rm eff} = (\sqrt{2}G_{F})^{1/2} \frac{m_{\ell i} C_{ij}}{\cos^2\beta}
  \bar{\ell}_{iR} \ell_{jL }\left[\sin(\alpha-\beta)H^{0} +\cos(\alpha-\beta)h^{0} - i A^{0} \right]+h.c.
  \label{eq:higgs-mediated}
 \end{eqnarray}
with $m_{\ell i}$ is the mass of the ith flavor lepton and $C_{ij}=\epsilon_{2} I_{ij}/(1+(\epsilon_1+\epsilon_2 I_{ii})\tan\beta)^2$.


 From Eq.~(\ref{eq:higgs-mediated}), we see that
the decays of $\tau\to \ell P$  only pick up the contributions
from the pseudoscalar boson $A^{0}$,
 while $\tau\to \ell S$ and $\tau \to \ell
PP$ are governed by both scalar bosons $H^0$ and $h^{0}$ due to the parity properties.
In our following analysis, we only concentrate on the
processes associated with the productions of $s\bar{s}$
 and $\mu^{+} \mu^{-}$ pairs to avoid small Higgs couplings.
  We choose the
decays of $\tau\to \ell X$ with $X=\mu^{+} \mu^{-}$, $\eta^{(\prime)}$,
$f_{0}(980)(\sigma(600))$ and $K^{+} K^{-}$ as the representative
modes.
For $\tau\to \ell \mu^{+} \mu^{-}$, the formalisms for the
decay rates dictated by  scalar and pseudoscalar bosons are given by
\begin{eqnarray}
\Gamma(\tau\to \ell \mu^{+} \mu^{-})\simeq c_{\ell} \frac{G^2_{F}
m^2_{\mu} m^{7}_{\tau}|C_{\tau \ell}|^2}{3\cdot 2^{9} \pi^3
\cos^{6}\beta} \left[\left(
\frac{cs}{m^2_{h}}-\frac{sc}{m^2_{H}}\right)^2
+\left(\frac{\sin\beta}{m^2_{A}}\right)^2 \right]\, ,
\label{eq:mumu}
\end{eqnarray}
where $c_{\ell}=3/2$ and 1 with $\ell=\mu$ and $e$,
$cs=\cos(\alpha-\beta)\sin\alpha$ and
$sc=\sin(\alpha-\beta)\cos\alpha$, respectively.
To study the production of $\eta^{(\prime)}$, we adopt the
quark-flavor
scheme, defined by \cite{flavor}
\begin{eqnarray}
\left( {\begin{array}{*{20}c}
   \eta   \\
   {\eta '}  \\
\end{array}} \right) = \left( {\begin{array}{*{20}c}
   {\cos \phi } & { - \sin \phi }  \\
   {\sin \phi } & {\cos \phi }  \\
\end{array}} \right)\left( {\begin{array}{*{20}c}
   {\eta _{q} }  \\
   {\eta _{s} }  \\
\end{array}} \right) \,,\label{eq:flavor}
\end{eqnarray}
where $\eta _{q}  = ( {u\bar u + d\bar d})/\sqrt{2}$ and $\eta_{s} =
s\bar s $.
 From $\langle 0| \bar q' \gamma_{\mu} \gamma_{5} q'|
\eta_{q'}(p)\rangle=f_{\eta_{q'}}p_{\mu}$, the mass of
$\eta_{q(s)}$ can be expressed by
$m_{qq}^2  = \frac{\sqrt 2}{f_q }\langle 0|m_u \bar u\gamma _5 u +
m_d \bar d\gamma _5 d| \eta_q \rangle$
 ($ m_{ss}^2  =
\frac{2}{f_s}\langle 0 |m_s \bar s\gamma _5 s| \eta _s \rangle$).
If we neglect the $\eta_{q}$ contribution due to small $m_{u,d}$,
the decay rates for
$\tau\to \ell \eta$ can be written as
\begin{eqnarray}
\Gamma(\tau\to \ell \eta)\simeq \frac{G^2_{F}m^{3}_{\tau}|C_{\tau
\ell}|^2}{64\pi}\tan^6\beta \left( \sin\phi f_{s}
\frac{m^2_{ss}}{m^2_{A}}\right)^2
\left(1-\frac{m^2_{\eta}}{m^2_{\tau}} \right)^2\,. \label{eq:eta}
\end{eqnarray}
Similarly, the rate for $\tau\to \ell \eta^{\prime}$ is given by
\begin{eqnarray}
\frac{\Gamma(\tau\to \ell \eta^{\prime})}{\Gamma(\tau\to \ell
\eta)}=\cot^2\phi
\left(\frac{1-m^2_{\eta^{\prime}}/m^2_{\tau}}{1-m^2_{\eta}/m^2_{\tau}}\right)^2\,
. \label{eq:rateetap}
\end{eqnarray}
 For $\tau\to \ell f_{0}(980)(\sigma(600))$ decays, although the quark contents of $f_{0}(980)$ and $\sigma(600)$ are
still uncertain,
we adopt two quark contents to
describe the states.
In terms of the notations in
Refs.~\cite{Chen,CCY}, the isoscalar states $f_{0}(980)$ and
$\sigma(600)$
 are described by
   $ |f_{0}(980) \rangle = \cos\theta |s\bar{s}\rangle + \sin\theta |
    n\bar{n}\rangle $ and
   $ |\sigma(600) \rangle = -\sin\theta |s\bar{s}\rangle + \cos\theta |
    n\bar{n}\rangle $
  with $n\bar{n}=(u\bar{u}+d\bar{d})/\sqrt{2}$ and $\theta$ being
  the mixing angle. The decay constants are defined as
  \begin{eqnarray}
   \langle f^{s}_{0}|\bar{s} s| 0\rangle &=& m_{f_0}
   \tilde{f}^{s}_{f_0}
          , \ \ \      \langle \sigma^{s}|\bar{s} s| 0\rangle =  m_{\sigma}
     \tilde{f}^{s}_{\sigma}\,,
  \end{eqnarray}
where $f^{s}_{0}$ and $\sigma^{s}$ represent the $s \bar{s}$
component in $f_{0}(980)$ and $\sigma(600)$, respectively.
As a result, the decay
rates of $\tau\to \ell f_{0}(980)$ are given by
\begin{eqnarray}
\Gamma(\tau\to \ell f_{0}(980))\simeq \frac{G^2_{F}
m^3_{\tau}|C_{\tau \ell}|^2}{16\pi \cos^6 \beta} \left(m_s m_{f_0}
\tilde{f}^{s}_{f_0} \cos\theta \right)^2 \left(
\frac{cs}{m^2_{h}}-\frac{sc}{m^2_{H}}\right)^2
\left(1-\frac{m^2_{f_0}}{m^2_{\tau}} \right)^2 \,. \label{eq:f0}
\end{eqnarray}
On the other hand,
the rates for $\tau\to \ell \sigma(600)$ can be obtained by
\begin{eqnarray}
\frac{\Gamma(\tau\to \ell \sigma(600))}{\Gamma(\tau\to \ell
f_{0}(980))}\simeq \left(
\frac{m_{\sigma}\tilde{f}^{s}_{\sigma}\tan\theta
}{m_{f_0}\tilde{f}^{s}_{f_0}}\right)^2
\left(\frac{1-m^2_{\sigma}/m^2_{\tau}}{1-m^2_{f_{0}}/m^2_{\tau}}\right)^2\,.
\label{eq:ratesigma}
\end{eqnarray}

For the three-body decays of $\tau\to \ell K^{+} K^{-}$, the
associated hadronic effects are much more complicated and unclear.
Nevertheless, the uncertainties could be fixed by the $B$ decays,
such as $B\to KKK$.
The related form factor
including resonant and nonresonant effects is defined by \cite{CCS}
\begin{eqnarray}
\langle K^{+}(p_1) K^{-}(p_2)|\bar{s} s|0 \rangle \equiv
f^{K^{+}K^{-}}_{s}(Q^2)=\sum_{S} {m_{S}\, \tilde{f}^{s}_{S} \,
g^{S\to KK} \over m^2_{S}-Q^2-im_{S} \Gamma_{S}} +f^{NR}_{s}\,,
\end{eqnarray}
where $S$ stands for the possible scalar meson state,
$m_{S} \tilde{f}^{s}_{S}=\langle S|\bar{s}
s |0\rangle $, $g^{S\to KK}$ denotes the
strong coupling for $S\to KK$, and
\begin{eqnarray}
  f^{NR}_{s}&=& \frac{v}{3}\left(3F^{1}_{NR}+2F^{2}_{NR}\right)
  + v\frac{\kappa}{Q^2} \left(\ln\frac{Q^2}{\Lambda^2}\right)^{-1}\,,
\nonumber\\
F^{1(2)}_{NR}&=&
\left(\frac{x^{1(2)}_{1}}{Q^2}+\frac{x^{1(2)}_{2}}{Q^4} \right)
\left( \ln \frac{Q^2}{\Lambda^2}\right)^{-1}\,,
\end{eqnarray}
with $v=(m^2_{K}-m^2_{\pi})/(m_s-m_d)$,
$x^{1}_{1}=-3.26$ GeV$^2$,
$x^{1}_{2}=5.02$ GeV$^2$, $x^{2}_{1}=0.47$ GeV$^2$ and
$x^{2}_{2}=0$.
 It is found that only
$f_{0}(980)$ and $f_{0}(1530)$ have the largest couplings to the $KK$
pair \cite{ANS}.
Note that in calculating $B\to KKK$ \cite{CCS},
the factorization approach in Ref. \cite{BBNS} has been used.
In our numerical estimations, we will only consider
these two scalar contributions.
 The differential decay rates as a function
of the invariant mass in the $KK$ system are given by
\begin{eqnarray}
\frac{d\Gamma(\tau\to \ell K^{+}K^{-})}{dQ^2}&\simeq &\frac{G^2_{F}
m^3_{\tau}|C_{\tau \ell}|^2}{2^{8}\pi^3 \cos^6\beta} \left( m_s
f^{K^+K^-}_{s}\right)^2 \left(
\frac{cs}{m^2_{h}}-\frac{sc}{m^2_{H}}\right)^2 \nonumber \\
&\times & \left( 1- \frac{Q^2}{m^2_{\tau}}\right)^2
\left(1-\frac{4m^2_{K}}{Q^2}\right)^{1/2}\,.
\end{eqnarray}
 From Eqs.~(\ref{eq:mumu}), (\ref{eq:eta}) and (\ref{eq:f0}),
it is interesting to see that
  the various decay rates mediated by
the Higgs bosons have the relationship
\begin{eqnarray}
\Gamma(\tau\to \ell \mu^{+} \mu^{-} )= \frac{c_{\ell} m^2_{\mu}
m^{4}_{\tau}}{3\cdot 2^5 \pi^2} \left[ \frac{\Gamma(\tau\to \ell
\eta)}{C_{\eta}} + \frac{\Gamma(\tau\to \ell
f_{0}(980))}{C_{f_0}}\right]\,, \label{eq:ell2mu}
\end{eqnarray}
where $C_{\eta}= (\sin^2\beta \sin\phi f_{s}
m^2_{ss}/2)^2(1-m^2_{\eta}/m^2_{\tau}))^2$ and $C_{f_0}=(m_{s}
m_{f_0} \tilde{f}^{s}_{f_0}
\cos\theta)^2(1-m^2_{f_0}/m^2_{\tau})^2$.

\begin{figure}[htbp]
\includegraphics*[width=4.5in]{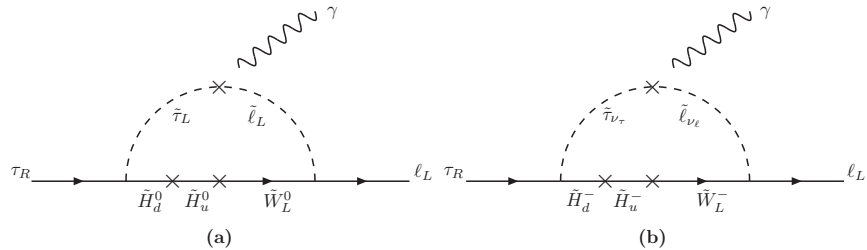}  \caption{The Feynman diagrams for $\tau\to\ell \gamma$ with a large $\tan\beta$.
The corsses represent the various mixing effects.}
 \label{fig:LFV}
\end{figure}

We now consider
the radiative modes of $\tau\to \ell \gamma$.
At the large $\tan\beta$ scenario, the dominant contributions to the decays are illustrated in Fig.~\ref{fig:LFV}.
To simplify the estimations,
we use the mass insertion method to formulate the decay
amplitudes.
The induced LFVs in the slepton mass matrix can
be approximately written as \cite{BM,HMTY,CKKL_PRD72}
\begin{eqnarray}
\left(\Delta m^{2}_{\tilde{L}}\right)_{ij}\simeq
-\frac{1}{(4\pi)^2} \left(6m^{2}_{0} Y^{\dagger}_{\nu} Y_{\nu}+2
A^{\dagger} A \right)_{ij} \ln\left(\frac{M_{U}}{M_{R}}\right)
\label{eq:smass}
\end{eqnarray}
where $m_{0}$, $Y_{\nu}$ and $A_{\nu}$ denote the typical initial
soft SUSY-breaking mass of the slepton, the neutrino Yukawa couplings
and the trilinear soft SUSY-breaking effects, respectively, at the
 unified scale of $M_{U}$.
 From Fig.~\ref{fig:LFV}
 and Eq. (\ref{eq:smass}),
  the effective interactions
for $\tau \to \ell \gamma$ are given by
\begin{eqnarray}
T&=& {G_{F}\over \sqrt{2}}\, e\, m_{\tau}\epsilon^{\mu*}(k)
\bar{\ell}(p-k) i\sigma_{\mu\nu}k^{\nu} A_{R} (1+\gamma_{5}) \tau(p)\,,
\label{eq:photon}
\end{eqnarray}
where
 \begin{eqnarray}
A_{R}&=& \frac{M_2 \mu}{(4\pi)^2} \frac{ m^2_{W}}{M^2_2} \tan\beta
(\Delta m^{2}_{\tilde{L}})_{\tau
\ell}\sum_{S=\tilde{\ell},\tilde{\nu}_{\ell}}  G_{S}\, ,
\label{eq:AR}
\end{eqnarray}
\begin{eqnarray}
G_{\tilde{\ell}}&=& -\frac{1-\tan^2\theta_{W}
}{m^2_{\tilde{\ell}_{L}}-m^2_{\tilde{\tau}_{L}}}\left[
\frac{f_{n}(x_{\tilde{\ell}_{L}})}{m^{2}_{\tilde{\ell}_{L}}} -
\frac{f_{n}(x_{\tilde{\tau}_{L}})}{m^{2}_{\tilde{\tau}_{L}}}\right],\
%
   G_{\tilde{\nu}_{\ell}}= \frac{4}{m^2_{\tilde{\nu}_{\ell}}-m^2_{\tilde{\nu}_{\tau}}}\left[
\frac{f_{c}(x_{\tilde{\nu}_{\ell}})}{m^{2}_{\tilde{\nu}_{\ell}}} -
\frac{f_{c}(x_{\tilde{\nu}_{\tau}})}{m^{2}_{\tilde{\nu}_{\tau}}}\right],\,
\nonumber \\
  f_{n}(x)&=& \frac{1}{(1-x)^3} \left(1-x^2+2x\ln x \right),\
%
  f_{c}(x)= -\frac{1}{2(1-x)^3} \left( 3-4x+x^2+2\ln x \right)\,,
  \label{eq:ar}
\end{eqnarray}
with $x_{S}=M^2_{2}/m^{2}_{S}$ \cite{HMTY}. Here,
we have set the masses of higgisions
and gauginos to be the same, denoted as $M_{2}$.
Subsequently, the decay rates of $\tau\to \ell \gamma$ are given by
\begin{eqnarray}
\Gamma(\tau \to \ell \gamma)=\frac{\alpha_{em}}{2}
G^{2}_{F}m^{5}_{\tau} |A_{R}|^2.
\end{eqnarray}
The diagrams in Fig.~\ref{fig:LFV} can also induce $\tau\to \ell\mu^{+}\mu^{-}$, $\tau\to \ell\phi$ and $\tau\to \ell K^{+}K^{-}$
when the photon is off-shell.
 From Eq.~(\ref{eq:photon}), it is easy to estimate the ratios
 of branching ratios (BRs) to be
\cite{HMTY}
\begin{eqnarray}
\label{eq:Rmu}
R_{\ell}^{\gamma}={{\rm BR}(\tau\to \ell X_{\gamma})_{\gamma}\over {\rm BR}(\tau\to \ell \gamma)}
=
O\left({\alpha_{em}\over \pi}\right)\sim 10^{-3}\,, \ (X_{\gamma}=\mu^{+}\mu^{-}\,,\ \phi\,,\ K^{+}K^{-})\,.
\end{eqnarray}
Note that it is impossible to produce modes with $X_{\gamma}$ being
a single pseudoscalar or scalar by the dipole operators in Eq.
(\ref{eq:photon}).
In our estimations for the modes with
$X_{\gamma}=\phi$ and $K^{+}K^{-}$, we have used the hadronic matrix elements defined
by
$\langle 0| \bar{q}\gamma^{\mu} q| \phi
\rangle= im_{\phi} f_{\phi} \epsilon^{*}_{\phi}(k)$ and
$ \langle 0 | \bar{q} \gamma^{\mu} q |
 K^{+}(p_1) K^{-}(p_2)\rangle  = (p^{\mu}_{1}-p^{\mu}_{2}) F^{K^{+} K^{-}}_{q}(Q^2)$, with the form factors given in Refs. \cite{fphi} and \cite{CCS},
 respectively.
 It is clear that from the current limits on
 $BR(\tau\to \ell\gamma)$, $BR(\tau\to \ell X_{\gamma})_{\gamma}$ are too small to
 be observed.
 We remark that other loop contributions to the decays,
 such as those from box diagrams, are expected to be small due to the light fermion final states.

For the numerical estimations on
$\tau \to \ell \gamma$ and $\tau\to \ell X$,
 we assume that $M_{1}\sim
M_{2}\sim m_{0} \sim \mu \sim m_{\tilde{\ell}}\sim m_{\tilde{\tau}}
$ to simplify our discussions.
Consequently, Eqs.~(\ref{eq:epsilon}) and (\ref{eq:AR}) become
 \begin{eqnarray}
  \epsilon_{1}& \approx & \frac{3\alpha_{em}}{4\pi \sin^2(2\theta_W)}, \ \ \
    \epsilon_{2}\approx \frac{\alpha_{em}}{16\pi }
    \left(\frac{1}{3\cos^2\theta_{W}}+\frac{1}{\sin^2\theta_{W}} \right)\, ,
    \nonumber \\
     A_{R}& \approx &\frac{1}{6(4\pi)^2} \frac{m^2_{W}}{m^2_{\tilde{\tau}}}
  \frac{(\Delta
m^{2}_{\tilde{L}})_{\tau \ell}}{m^2_{0}}
\tan\beta(1+\tan^2\theta_{W}),
 \end{eqnarray}
 respectively.
If we regard $A^{\dagger}A$ in Eq.~(\ref{eq:smass}) as
$(A^{\dagger}A)_{\tau \ell} \sim m^2_{0}(Y^{\dagger}_{\nu}
Y_{\nu})_{\tau \ell}= m^2_{0}O(1)$,
we get $(\Delta m^{2}_{\tilde{L}})_{\tau \ell}/m^{2}_{0}\sim -8/(4\pi)^{2}
\ln(M_{u}/M_{R})$. Thus,
we find
that $C_{\tau \ell}$ are insensitive to the SUSY breaking scale
and the decays of $\tau\to \ell \gamma$ and
$\tau\to \ell X$ are only sensitive to the masses of the
slepton and Higgs bosons, respectively. In calculating the numerical
values,
we set $G_{U(R)}=10^{19}(10^{14})$ GeV and
$\tan\beta=60$. Other parameters
in various modes are taken to be as follows: $\phi=39^{\circ}$,
$f_{s}=0.17$ GeV and $m_{ss}=0.69$ GeV for $\tau\to \ell
\eta^{(\prime)}$ \cite{flavor}; $\theta=30^{\circ}$, $m_{s}=0.15$
GeV and $\tilde{f}^{s}_{\sigma}\sim \tilde{f}^{s}_{f_0}= 0.33$ GeV
\cite{CCY} for $\tau\to \ell (f_{0}(980),\sigma(600))$; $v=2.87$
GeV, $\kappa=-10.4$ GeV$^{4}$, $\tilde{f}^{s}_{f_0(1530)}\sim
\tilde{f}_{f_0(980)}=0.33$ GeV, $g^{f_0(980)\to KK}=1.50$ GeV,
$g^{f_0(1530)\to KK}=3.18$ GeV \cite{CCS}, $\Gamma_{f_0(980)}=80$
MeV and $\Gamma_{f_0(1530)}=1.16$ GeV \cite{ANS} for $\tau\to \ell
K^{+} K^{-}$. For simplicity, we do not distinguish the
difference between $(Y^{\dagger}_{\nu} Y_{\nu})_{\tau e}$ and
$(Y^{\dagger}_{\nu} Y_{\nu})_{\tau \mu}$, $i.e.$, $(\Delta
m^{2}_{\tilde{L}})_{\tau e}=(\Delta m^{2}_{\tilde{L}})_{\tau \mu}$.

\begin{figure}[htbp]
\includegraphics*[width=2.5 in]{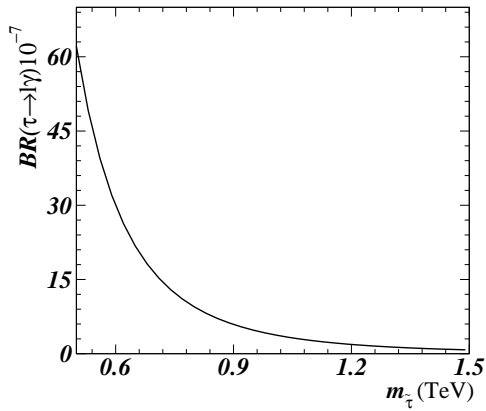}
\caption{Branching ratios (in units of $10^{-7}$ ) for $\tau\to \ell
\gamma$ as a function of the stau mass. }
 \label{fig:tauellga}
\end{figure}

\begin{figure}[htbp]
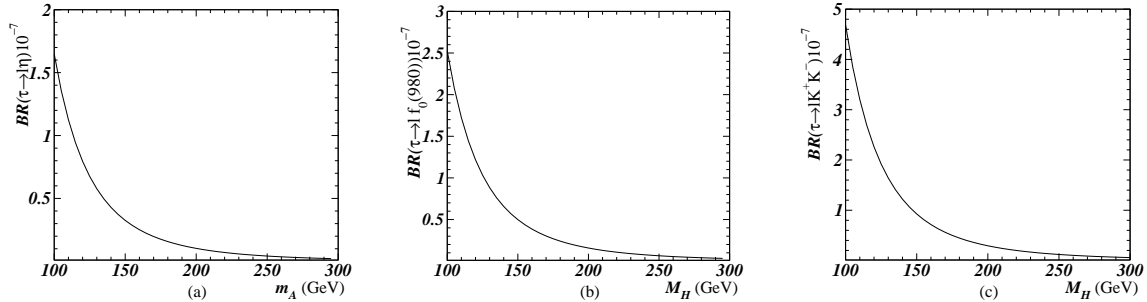

\includegraphics*[width=1.8 in]{tauelleta}\hspace{0.5cm}
\includegraphics*[width=1.8 in]{tauellf0} \hspace{0.5cm}
\includegraphics*[width=1.8 in]{tauellkk}
\caption{Branching ratios (in units of $10^{-7}$) for (a) $\tau\to
\ell \eta$ and  (b)[(c)] $\tau\to \ell f_{0}(980)[K^{+} K^{-}]$ as
functions of the pseudoscalar and scalar Higgs masses, respectively. }
 \label{fig:tauelleta}
\end{figure}

In Fig.~\ref{fig:tauellga},
 we present the
BRs for $\tau\to \ell \gamma$ as a function of the slepton mass.
%
%
In comparison with the BELLE and BABAR results of
$BR(\tau\to \mu \gamma) < 3.1 \times 10^{-7}$ \cite{belle-tau}
and $0.68 \times 10^{-7}$ \cite{babar-tau},
we see clearly that $m_{\tilde{\tau}}> 1$~TeV is favorable. The BRs
of $\tau\to \ell \eta$ as a function of the pseudoscalar mass are displayed
in Fig.~\ref{fig:tauelleta}(a). From Eq.~(\ref{eq:rateetap}), we have
$BR(\tau\to \ell \eta^{\prime})=0.93 BR(\tau\to \ell \eta)$. The BRs
of $\tau\to \ell f_0(980)$ and $\tau\to \ell K^{+}K^{-}$ as a
function of $M_{H}=(cs/m^2_h-sc/m^2_{H})^{-1/2}$ are shown in
Figs.~\ref{fig:tauelleta}(b) and \ref{fig:tauelleta}(c), respectively.
In terms of Eq.~(\ref{eq:ratesigma}), we get $BR(\tau\to \ell
\sigma(600))=0.2BR(\tau\to \ell f_0(980))$. In addition, from
Eq.~(\ref{eq:ell2mu}),
we obtain that $BR(\tau\to \ell \mu^{+}
\mu^{-})\simeq 0.33\left[BR(\tau\to \ell \eta)+1.6 BR(\tau\to \ell
f_{0}(980) \right]$. Clearly,
all $\tau\to\ell X$ modes except $\tau\to \ell \sigma(600)$
are suitable to search for the LFV. Finally, it is worth mentioning
that if we take the decoupling limit, i.e. $m_{H}\approx m_{A}$ and
$\alpha\to \beta-\pi/2$ \cite{Higgs}, leading to $M_{H}=m_{H}$, we
get $\Gamma(\tau\to\ell f_{0}(980)): \Gamma(\tau\to \ell \mu^{+}
\mu^{-}) :\Gamma(\tau\to \ell \eta)\approx 1.3\, :\, 0.36c_{\ell} \,
:\, 1$.

In summary, we have studied the lepton flavor violating $\tau$ decays
through the Higgs-mediated mechanism with the nonholomorphic terms from
the couplings between the Higgs bosons and leptons at the large $\tan\beta$.
By assuming that all masses associated with SUSY-breaking are the same,
we have demonstrated that  BRs of $\tau\to \ell \gamma$ only depend on the
stau mass. In the Higgs-mediated mechanism,
 we have shown that the BRs of the
new proposed decays  of $\tau\to \ell f_0(980)$ and $\tau\to \ell K^{+}
K^{-}$ arising from the scalar exchanges can be as large as the upper limits $O(10^{-7})$ of the current data and, moreover, they can be larger than those of $\tau\to \ell \eta$ from pseudoscalar exchanges.
We have also pointed out that
 $\tau\to \ell \mu^{+} \mu^{-}$
 are related with $\tau\to \ell \eta$ and $\tau\to \ell f_0(980)$.
 It is clear that future experimental searches for the LFV in the
 leptonic and  semileptonic tau flavor violating decays are important for
 us to identify the Higgs-mediated mechanism.\\

\noindent
{\bf Acknowledgments}

This work is supported in part by the National Science Council of
R.O.C. under Grant \#s:NSC-94-2112-M-006-009 and
NSC-94-2112-M-007-004.


\begin{thebibliography}{99}

\bibitem{atom}SuperKamiokande Collaboration, S. Fukuda {\it et al.}, Phys. Rev. Lett. {\bf 85}, 3999
(2000).

\bibitem{solar} SuperKamiokande Collaboration, S. Fukuda {\it et al.}, Phys.
Rev. Lett. {\bf 86}, 5656 (2001);  SNO Collaboration, Q.R. Ahmad
{\it et al.}, Phys. Rev. Lett. {\bf 87}, 071301 (2001); {\bf 89},
011301 (2002).

\bibitem{seesaw}M. Gell-Mann, P. Ramond and R. Slansky, in Supergravity, edited by P. Van Nieuwenhuizen
and D.Z. Freedman (North-Holland, New-York, 1979); T. Yanagida, in
Proceedings of the Workshop on Unified Theory and Baryon Number in
the Universe, edited by O. Sawada and A. Sugamoto (KEK, Tsukuba,
1979); R.N. Mohapatra and G. Senjanovi$\rm \acute{c}$, Phys. Rev.
Lett. {\bf 44}, 912 (1980).


\bibitem{BM}F. Borzumati and A. Masiero, Phys. Rev. Lett. {\bf 57},
961 (1986).

\bibitem{HMTY} J. Hisano {\it et al.}, Phys. Rev. D{\bf 53}, 2442
(1996); J. Hisano and D. Nomura, Phys. Rev. D{\bf 59}, 116005
(1999).

\bibitem{BK_PRL}K.S. Babu and C. Kolda, Phys. Rev. Lett. {\bf 89},
241802 (2002).


\bibitem{LFV1} M. Sher, Phys. Rev. D{\bf 66}, 057301 (2002); D. Black
{\it et al.}, Phys. Rev. D{\bf 66}, 053002 (2002).

\bibitem{LFV2} A. Dedes, J.R. Ellis and M. Raidal, Phys. Lett. B{\bf 549}, 159
(2002); A. Brignole and A. Rossi, Nucl. Phys. B{\bf 701}, 3 (2004).

\bibitem{MORE}
A.~Ilakovac, B.~A.~Kniehl and A.~Pilaftsis,
  Phys.\ Rev.\ D {\bf 52}, 3993 (1995);
  A.~Ilakovac,
  Phys.\ Rev.\ D {\bf 54}, 5653 (1996);
  A.~Atre, V.~Barger and T.~Han,
  Phys.\ Rev.\ D {\bf 71}, 113014 (2005);
T.~Fukuyama, A.~Ilakovac and T.~Kikuchi,
  arXiv:hep-ph/0506295;
   V.~Cirigliano and B.~Grinstein,
  arXiv:hep-ph/0601111.

\bibitem{mu-e} R. Kitano {\it et al.},
Phys. Lett. B{\bf 575}, 300 (2003).

\bibitem{mu-ePara}
A.~Masiero, P.~Paradisi and R.~Petronzio,
  arXiv:hep-ph/0511289.

\bibitem{mu-tau} M. Sher and I. Turan, Phys. Rev. D{\bf 69}, 017302
(2004).

\bibitem{LFV_collider}  A. Brignole and A. Rossi, Phys. Lett. B{\bf 566}, 217
(2003); S. Kanemura {\it et al.},
Phys. Lett. B{\bf 599}, 83 (2004); E. Arganda {\it et al.},
Phys. Rev. D {\bf 71}, 035011 (2005).


\bibitem{Hamzaoui:1998nu}
  C.~Hamzaoui, M.~Pospelov and M.~Toharia,
  %
  Phys.\ Rev.\ D {\bf 59} (1999) 095005.
\bibitem{qf1}T. Banks, Nucl. Phys. B{\bf 303}, 172 (1988); E. Ma, Phys.
Rev. D{\bf 39}, 1922 (1989); R. Hempfling, Phys. Rev. D{\bf 49},
6168 (1994);
K. S. Babu, B. Dutta, and R. N.
Mohapatra, Phys. Rev. D{\bf 60}, 095004 (1999);
L.J. Hall, R. Rattazzi, and U. Sarid, Phys. Rev. D{\bf 50}, 7048
(1994);
F.~Borzumati, G.~R.~Farrar, N.~Polonsky and S.~D.~Thomas,
  Nucl.\ Phys.\ B {\bf 555}, 53 (1999).
\bibitem{qf2}
A.~Dedes and A.~Pilaftsis,
  %
  Phys.\ Rev.\ D {\bf 67}, 015012 (2003),
A.~J.~Buras, P.~H.~Chankowski, J.~Rosiek and L.~Slawianowska,
  %
  Phys.\ Lett.\ B {\bf 546}, 96 (2002).


\bibitem{IR} G. Isidori and A. Retico, JHEP {\bf 0111}, 001 (2001); JHEP {\bf 0209}, 063
(2002).

\bibitem{Parry}J.K. Parry, arXiv:hep-ph/0510305.



\bibitem{PDG04} Particle Data Group, S. Eidelman {\it et al.}, Phys.
Lett. B{\bf 592}, 1 (2004).

\bibitem{belle-tau} Belle Collaboration, K. Abe {\it et al.}, Phys. Rev. Lett. {\bf 92}, 171802
(2004); K. Hayasaka {\it et al.}, Phys. Lett. B{\bf 613}, 20 (2005).

\bibitem{babar-tau} Babar Collaboration, B. Aubert {\it et al.},
Phys. Rev. Lett. {\bf 95}, 041802 (2005); Phys. Rev. Lett. {\bf 96},
041801 (2006).


\bibitem{belle-tau-hh} Belle Collaboration, Y. Enari {\it et al.}, Phys. Lett. B{\bf 622}, 218 (2005);
 Y. Yusa {\it et al.}, arXive:hep-ex/0603036.

\bibitem{LYZ} W.J. Li, Y.D. Yang, X.D. Zhang,
Phys. Rev. D{\bf 73}, 073005
(2006).

\bibitem{Higgs} J. F. Gunion {\it et al.}, "The Higgs Hunter's Guide" (Addison-Wesley, Reading, MA 1990);
Erratum: arXiv:hep-ph/9302272.

\bibitem{flavor}T. Feldmann, P. Kroll, B. Stech, Phys. Rev. D{\bf 58},
114006(1998).

\bibitem{Chen} C.H. Chen, Phys. Rev. D{\bf 67}, 014012 (2003); 094011
(2003).

\bibitem{CCY}H.Y. Cheng, C.K. Chua and K.C. Yang, Phys. Rev. D{\bf 73}, 014017 (2006).


\bibitem{CCS} H.Y. Cheng, C.K. Chua and A. Soni, Phys. Rev. D{\bf 72}, 094003 (2005).


\bibitem{ANS} V.V. Anisovich, V.A. Nikonov and A.V. Sarantsev, Phys. Atom. Nucl. {\bf 65}, 1545
(2002).


\bibitem{BBNS}
  M.~Beneke, G.~Buchalla, M.~Neubert and C.~T.~Sachrajda,
  Phys.\ Rev.\ Lett.\  {\bf 83}, 1914 (1999); Nucl.\ Phys.\ B {\bf 591}, 313 (2000).

\bibitem{CKKL_PRD72}
J.A. Casas and A. Ibarra, Nucl. Phys. {\bf B618}, 171 (2001);
K. Cheung {\it et al.}, Phys. Rev. D{\bf 72}, 036003
(2005).


\bibitem{fphi}
P. Ball, V.M. Braun, Y. Koike, K. Tanaka, Nucl.Phys. B529, 323 (1998).




\end{thebibliography}
\end{document}